\documentclass[a4paper]{jpconf}
\usepackage{graphicx}
\usepackage{color}
\begin{document}
\title{${}^{7}$Li($d$,$p$)${}^{8}$Li  transfer reaction  in \\ the NCSM/RGM approach}

\author{F Raimondi$^{1,2,a}$, G Hupin$^{3,4,5,b}$, P Navr\'atil$^{2,c}$ and S Quaglioni$^{4,d}$}
\address{$^{1}$Department of Physics, Faculty of Engineering and Physical Sciences, University of Surrey, Guildford GU2 7XH, United Kingdom. \\ $^{2}$TRIUMF, 4004 Wesbrook Mall, Vancouver BC, V6T 2A3, Canada. \\
$^{3}$CEA, DAM, DIF, F-91297 Arpajon, France. \\
$^{4}$Lawrence Livermore National Laboratory, P.O Box 808, L-414, Livermore, California 94551, USA \\
$^{5}$Institut de Physique Nucl\'eaire, IN2P3-CNRS, Universit\'e Paris-Sud, F-91406 Orsay  Cedex, France. }
 \ead{\\ $^{a}$f.raimondi@surrey.ac.uk \\ $^{b}$hupin@ipno.in2p3.fr \\ $^{c}$navratil@triumf.ca \\ $^{d}$quaglioni1@llnl.gov}

\begin{abstract}
Recently,  we  applied an \textit{ab initio} method, the no-core shell model combined with the resonating group method, to the transfer reactions with light p-shell nuclei as targets and deuteron as the projectile. In particular, we studied the elastic scattering of deuterium on $^7$Li and the ${}^{7}$Li($d$,$p$)${}^{8}$Li  transfer reaction starting  from a realistic two-nucleon interaction. 
In this contribution, we review of our main results on the ${}^{7}$Li($d$,$p$)${}^{8}$Li  transfer reaction, and we extend the study of the relevant reaction channels, by showing the dominant resonant phase shifts of the scattering matrix. We assess also the impact of the polarization effects of the deuteron below the breakup on the positive-parity resonant states in  the reaction. For this purpose, we perform an analysis of the convergence trend of the phase and eigenphase shifts, with respect to the number of deuteron pseudostates included in the model space.  
\end{abstract}

\section{\label{intro}Introduction}
The stripping of a proton ($p$) or a neutron ($n$) from the deuteron ($d$)  projectile is the simplest nuclear transfer reaction. The deuteron is the  first stable isotope synthesized after the Big Bang; henceforth, $(d,p)$ and $(d,n)$ are among the key reactions in the formation of the light elements in the primordial nucleosynthesis. Because of its simplicity, the transfer of a nucleon, stripped from the deuteron, into a target nucleus
is an efficient process implemented in the experimental studies of the ground-state (g.s.) and excited states of nuclei, of their energies, spin, and parity properties~\cite{Jones2013}. 

The aim of an \textit{ab initio} description of the deuteron-induced transfer reaction is to connect the dynamics of the scattering to the first principles, which in low-energy nuclear reaction are the nucleons, considered as the relevant degrees of freedom, interacting though realistic forces.
Such a description has been successfully carried out for light target nuclei  in the framework of two  \textit{ab initio} many-body methods: the no-core shell model~\cite{Navratil2009} combined with the resonating group method (NCSM/RGM)~\cite{Quaglioni2009}, and the no-core shell model with continuum (NCSMC)~\cite{Baroni2013a, Baroni2013b}, which are both suited for realistic nucleon-nucleon (NN) and three-nucleon (3N)  forces. The former is a cluster approach applied to the $A$-nucleon wave function, which is partitioned in a  $a$ nucleons  projectile and a ($A$-$a$) nucleons target, allowing for the description of scattering states. The latter is built by treating on the same footing the cluster states of the NCSM/RGM method, and the $A$-nucleons wave-functions computed in the NCSM approach. The NCSM/RGM and NCSMC approaches have been first applied to the elastic  ($d$-$\alpha$)~\cite{Navratil2011, Hupin2015} and $^3$H($d,n$)$^4$He and $^3$He($d,p$)$^4$He transfer reactions~\cite{Quaglioni2012, Navratil2012}. 

In a recent work~\cite{PhysRevC.93.054606}, we extended the scope of the NCSM/RGM description of the deuteron-induced transfer reactions to $p$-shell nuclei.  We studied the $^7$Li($d$,$p$)$^8$Li transfer reaction, whose excitation function below the deuteron-breakup energy is adopted as a calibration tool for  the measurement of the radiative proton capture on $^7$Be~\cite{Adelberger1998}. We could then compute the energy, spin and parity of the resonances above the $d$+$^7$Li threshold in the $^9$Be spectrum. Even in the energy range below the deuteron breakup, the internal excitations of the deuteron (i.e. polarization and virtual breakup) are expected to influence the shape of the cross section. Their correct description  poses a challenge to the approaches that do not include explicitly the breakup of the deuteron. In the NCSM/RGM, the  formalism for the description of the three-body continuum has been fully worked out with the introduction of the three-cluster  wavefunction ansatz~\cite{PhysRevLett.113.032503}, but it is not yet implemented in the transfer reaction calculations. In the present approach, the virtual breakup of the two-nucleon projectile is approximated by discretizing the continuum, obtained by considering the positive-energy states (pseudostates) of the NCSM spectrum. 

In this contribution, we revisit the main equations of the NCSM/RGM formalism, and our results for the $^7$Li($d$,$p$)$^8$Li transfer reaction of Ref.~\cite{PhysRevC.93.054606}, then we discuss the binary-cluster channels that contribute to the resonances above the $p$+$^8$Li threshold.  Moreover, we assess our treatment of the deuteron continuum, via an analysis of the convergence of the resonant phase and eigenphase shifts  in terms of the number of deuteron pseudostates included in the model space basis.

\section{\label{formalism}Basic equations of the NCSM/RGM formalism}
A general presentation of the equations of the NCSM/RGM formalism can be found in Ref.~\cite{Quaglioni2009}. The idea of the cluster wave function, typical of the RGM approach, is exploited by building the following expansion over the antisymmetrized binary-cluster channel states $|\Phi^{J^\pi T}_{\nu r}\rangle$,
\begin{eqnarray}
|\Psi^{J^\pi T}\rangle &=&  \sum_{\nu} \int dr \,r^2\frac{g^{J^\pi T}_{\nu}(r)}{r}\,\hat{\mathcal A}_{\nu}\,|\Phi^{J^\pi T}_{\nu r}\rangle\ 
\label{eq:trial},
\end{eqnarray}
with the coefficient of the expansion $g^{J^\pi T}_{\nu}(r)$ being the unknown amplitudes of the relative motion between the clusters.
The quantum numbers $J$, $\pi$, and $T$, are the total angular momentum, parity and isospin.

Due to the introduction of the auxiliary variable $r$ in Eq.~(\ref{eq:trial}), formally distinct from  the inter-cluster relative coordinate $\vec r_{A-a,a}=r_{A-a,a}\hat r_{A-a,a}$, the antisymmetrization operator $\hat{\mathcal A}_{\nu}$ acts only on the channel states. For the case of the binary cluster, the operator $\hat{\mathcal A}_{\nu}$ acts on the product state,
 \begin{eqnarray}
|\Phi^{J^\pi T}_{\nu r}\rangle &=& \Big [ \big ( \left|A_t\, \alpha_t I_t^{\,\pi_t} T_t\right\rangle \left |A_p\,\alpha_p I_p^{\,\pi_p} T_p\right\rangle\big ) ^{(s T)}\,Y_{\ell}\left(\hat r_{A-a,a}\right)\Big ]^{(J^\pi T)}\,\frac{\delta(r-r_{A-a,a})}{rr_{A-a,a}},
\label{basis}
\end{eqnarray}
where $\left|A_{t(p)}\, \alpha_{t(p)} I_{t(p)}^{\,\pi_{t(p)}} T_{t(p)}\right\rangle$ are translational-invariant eigenstates of the target (projectile). 
Each channel is identified by a index $\nu=\{A_t  \alpha_t I_t^{\pi_t}T_t; A_p \alpha_p I_p^{\pi_p}T_p; s\ell\}$ collecting the total spin $s$, the  relative orbital angular momentum $\ell$, and the quantum numbers of each cluster: angular momentum $I$, parity ${\pi}$, isospin $T$ and energy $\alpha$.

The target and projectile states in the binary-cluster channel are in turn obtained from the intrinsic Hamiltonian $\hat{H}= \hat{T}_{int}+\hat{V}$, composed by the internal kinetic energy $\hat{T}_{int}$ and nuclear interaction $\hat{V}$, by solving the following eigenvalue problem,
\begin{equation}
\hat{H} \left|A\, \alpha I^{\,\pi} T\right\rangle = E_\alpha \left|A\, \alpha I^{\,\pi} T\right\rangle\,.\label{NCSMeq}
\end{equation}
In the NCSM approach~\cite{Navratil2009} the eigenstates of Eq.~(\ref{NCSMeq}) are obtained by diagonalizing $\hat{H}$ in a model space spanned by a complete harmonic oscillator (HO) basis. The size of the basis is fixed by the maximum number  $N_{\rm max}$  of HO quanta, while the same HO frequency $\Omega$ is used for both clusters in the binary-cluster channel.

Note that the RGM cluster ansatz of Eq.~(\ref{eq:trial}) is linked to the NCSM method consistently, that is the same microscopic Hamiltonian of Eq.~(\ref{NCSMeq}) is used. The basis states $\hat{\mathcal A}_{\nu}\,|\Phi^{J^\pi T}_{\nu r}\rangle$ of Eq.~(\ref{eq:trial}) give rise to a coupled-channel set of equations for the unknowns $g^{J^\pi T}_{\nu}(r)$, 
\begin{equation}
\sum_{\nu}\int dr \,r^2\left[{\mathcal H}^{J^\pi T}_{\nu^\prime\nu}(r^\prime, r)-E\,{\mathcal N}^{J^\pi T}_{\nu^\prime\nu}(r^\prime,r)\right] \frac{g^{J^\pi T}_\nu(r)}{r} = 0\,,\label{RGMeq}
\end{equation}
with the norm and Hamiltonian kernel given by,
\begin{eqnarray}
{\mathcal N}^{J^\pi T}_{\nu^\prime\nu}(r^\prime, r) &=& \left\langle\Phi^{J^\pi T}_{\nu^\prime r^\prime}\right|\hat{\mathcal A}_{\nu^\prime}\hat{\mathcal A}_{\nu}\left|\Phi^{J^\pi T}_{\nu r}\right\rangle\,,\label{N-kernel}
\end{eqnarray}
 and
\begin{eqnarray}
{\mathcal H}^{J^\pi T}_{\nu^\prime\nu}(r^\prime, r) &=& \left\langle\Phi^{J^\pi T}_{\nu^\prime r^\prime}\right|\hat{\mathcal A}_{\nu^\prime}\hat{H}\hat{\mathcal A}_{\nu}\left|\Phi^{J^\pi T}_{\nu r}\right\rangle\,,\label{H-kernel}
\end{eqnarray}
respectively.

In our implementation of the deuteron-induced reactions,  two different mass partitions are taken into account in the asymptotic states: the one with the deuteron ($A$-$2$,$2$) and the one with the scattered nucleon ($A$-$1$,$1$). As a consequence, two types of kernels appear in the matrices~(\ref{N-kernel}) and~(\ref{H-kernel}). The diagonal kernels correspond to the elastic channel of the reaction, where the deuteron is present as a scattered particle in the asymptotic state or as a cluster in the intermediate composite nucleus, that is
\begin{eqnarray}
\mathcal{H}_{\nu'\nu}^{J^\pi T}(r',r)&=&
\left<\Phi_{\nu' r'}^{J^\pi T}\right|\hat{\mathcal{A}}_{(A-2,2)}H\hat{\mathcal{A}}_{(A-2,2)}
\left|\Phi_{\nu r}^{J^\pi T}\right>
 =\left<\Phi_{\nu' r'}^{J^\pi T}\right|H\hat{\mathcal{A}}^2_{(A-2,2)}
\left|\Phi_{\nu r}^{J^\pi T}\right> \nonumber \\
&=&\left[{T}_{\rm rel}(r')+\bar{V}_C(r')+E_{\alpha_1'}^{I_1'T_1'} +E_{\alpha_2'}^{I_2'T_2'}\right]
\mathcal{N}_{\nu'\nu}^{J^\pi T}(r', r)+\mathcal{V}^{J^\pi T}_{\nu' \nu}(r',r),
\end{eqnarray}
where ${T}_{\rm rel}(r')$ and $\bar{V}_C(r')$ are the relative kinetic energy 
and average Coulomb interaction, respectively; $E_{\alpha_1'}^{I_1'T_1'}$ and $E_{\alpha_2'}^{I_2'T_2'}$ are NCSM energy eigenvalues for the two clusters, and $\mathcal{V}^{J^\pi T}_{\nu' \nu}$ is the potential kernel (see Eq.~(B2) of Ref.~\cite{PhysRevC.93.054606}).
The off-diagonal kernels couple different mass partitions and correspond to the transfer process: the complete list of the expressions of the diagonal and coupling kernels, both for the norm and the Hamiltonian, can be found in Refs.~\cite{Navratil2011},~\cite{Quaglioni2012} and~\cite{PhysRevC.93.054606}. Once the norm and Hamiltonian kernels have been computed, the set of coupled integral-differential equations in~(\ref{RGMeq}) is solved on a Lagrange mesh within the microscopic R-matrix method~\cite{Baye1986}.

\section{\label{resul} Results on $^7$Li$(d,p)^8$Li transfer reaction and $^9$Be energy spectrum}

For the description of the $^9$Be spectrum above the $d$+$^7$Li threshold,   
and of the  $^7$Li$(d,p)^8$Li transfer reaction, we start from specific choices of the nuclear interaction and model space:
\begin{itemize}
\item \emph{Interaction}: The chiral N$^3$LO NN potential of Ref.~\cite{Entem2003},
evolved through a similarity renormalization group (SRG) transformation with evolution parameter $\Lambda$=2.02 fm$^{-1}$. We do not include 3N forces.
\item \emph{HO basis}: The nuclear wave function is expanded in the HO basis, with frequency of $\hbar\Omega=20$ MeV and two truncations corresponding to a total number of excitations above the $2\hbar\Omega$ minimum-energy configuration of $N_{\rm max}=6$ and 8. To match the corresponding absolute number of HO  quanta, we described the deuteron in  $N_{\rm max}=8$, and 10 model spaces, respectively.
\item \emph{Model space}:  Two $^7$Li states ($\frac{3}{2}^-$ g.s. and $\frac{1}{2}^-$ first excited state) and four $^8$Li states ($2^+$ g.s. and $1^+$, $3^+$, $0^+$ excited states)  are taken as the clusters in the binary products of Eq.~(\ref{basis}). For the deuteron, we included the g.s. and up to 4 pseudostates in the $^{3}S_1$-$^{3}D_1$ channel.   Tables~\ref{tab:tableNCSMstate} and~\ref{tab:d_pseudo} give the energies of $^7$Li and $^8$Li states, and deuteron g.s. and pseudostates, respectively. 
\end{itemize}

\begin{table}[h]
\caption{\label{tab:tableNCSMstate}%
Ground-state and excitation energies of $^7$Li and $^8$Li calculated within the NCSM with $N_{\rm max}$= 6, 8 and 10 in the HO basis and HO frequency $\hbar\Omega$=20 MeV, compared to the experiment. 
The values in 
the last column have been adjusted in order to reproduce the Q-value of the ${}^{7}$Li($d$,$p$)$^{8}$Li reaction, as explained in Section~\ref{sec:cross_sec}.
}
\begin{center}
\lineup
\begin{tabular}{llccccc}
\br  
\textrm{Nucleus} & \textrm{State} &\multicolumn{5}{c}{E (MeV)} \cr
& \textrm{ J$^\pi$ } & \multicolumn{3}{c}{\textrm{$N_{\rm max}$ }}&\multicolumn{1}{c}{\textrm{Exp  }}&\multicolumn{1}{c}{\textrm{Threshold  }}\cr
& & 6 & 8 & 10 & & \multicolumn{1}{c}{\textrm{${}^{7}$Li($d$,$p$)$^{8}$Li}} \cr
\mr
 $^7$Li & $\frac{3}{2}^-$ & -36.20 & -38.01 & -38.94 & -39.25  &-38.01\cr
            & $\frac{1}{2}^-$ & -35.80 &  -37.64 & -38.60  & -38.77  & -37.53 \cr
\mr
$^8$Li & $2^+$ & -37.60 & -39.66& -40.75  & -41.28 & -40.04\cr
            & $1^+$ & -36.36 & -38.47& -39.63 & -40.30 & -39.06\cr
           & $3^+$ & -34.76 & -36.78 & -37.86 &  -39.02 &  -37.78 \cr
           & $0^+$& -33.75 & -36.16 & -37.56 &   & -36.83\cr
 \br            
\end{tabular}
\end{center}
\end{table}

\begin{table}[h]
\caption{\label{tab:d_pseudo}%
Ground-state and pseudostate energies of the deuteron calculated within the NCSM, with $N_{\rm max}= 8$, 10 and 12 basis space, and HO frequency $\hbar\Omega$=20 MeV. 
In the calculations we included up to 4 pseudostates in the $^{3}S_1$-$^{3}D_1$ channel. }
\begin{center}
\lineup
\begin{tabular}{lccc}
\br  
&
\multicolumn{3}{c}{E (MeV)}\cr
&  \multicolumn{1}{c}{\textrm{$N_{\rm max}$= 8  }}&
\multicolumn{1}{c}{\textrm{$N_{\rm max}$= 10  }}&
\multicolumn{1}{c}{\textrm{$N_{\rm max}$= 12  }}\cr
\mr
 g.s. & -1.96 & -2.12  & -2.13\cr
 1$^*$     & 9.91 &  8.36 & 6.93 \cr
 2$^*$      & 15.22 &  12.82 & 11.06 \cr
 3$^*$      & 33.24 &  26.6  & 22. 80\cr
 4$^*$      & 40.20 &   33.23 & 28.45\cr    
 \br        
\end{tabular}
\end{center}
\end{table}

\subsection{\label{sec:phase}$^7$Li$(d,p)^8$Li scattering (eigen)phase shifts}

A first insight on the dynamics of $^7$Li$(d,p)^8$Li transfer reaction can be obtained from the different ($J^\pi T$) components of the scattering matrix. The different binary-cluster states of Eq.~(\ref{eq:trial}), having good angular momentum, parity and isospin, contribute to different partial waves in the scattering matrix. The impact of different partial waves can be inferred from the eigenphase shifts, while the importance of a specific binary-cluster channel state within a partial wave is given by the phase shifts.

\begin{figure}[h]
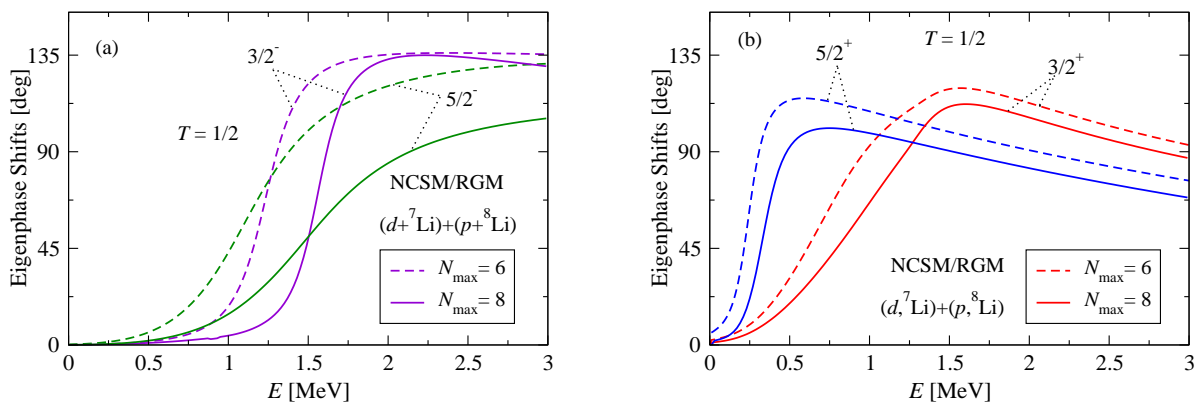

\begin{minipage}{17pc}
\includegraphics[width=17pc]{eigenphase_shift_p8Li_d7Li_srg-n3lo2p02_20_Nmax_6_8_4st_2st_RGM_MAIN_smooth_MINUS2.eps}
\end{minipage}\hspace{3pc}%
\begin{minipage}{17pc}
\includegraphics[width=17pc]{eigenphase_shift_p8Li_d7Li_srg-n3lo2p02_20_Nmax_6_8_4st_2st_RGM_MAIN_smooth_PLUS2.eps}
\end{minipage} 
\caption{\label{eigen_minus_plus}Calculated (a) negative and (b) positive-parity  eigenphase shifts within the coupled ($d,^7$Li)+($p,^8$Li) NCSM/RGM basis, as a function of the relative kinetic energy in the c.m.\ frame and with respect to the $p$+$^8$Li threshold. The SRG-N$^3$LO NN potential with $\Lambda$=2.02 fm$^{-1}$, and the HO frequency of $\hbar\Omega$=20 MeV were used.}
\end{figure}

 In Fig.~\ref{eigen_minus_plus}(a) and~(b)
we show a selection of the computed $T=\frac{1}{2}$ eigenphase shifts for negative- and 
positive-parity states, respectively. These partial waves are  responsible for  the shape and strength of the $^7$Li$(d,p)^8$Li cross section, as shown in Section~\ref{sec:cross_sec}.

\begin{figure}[h]
\begin{minipage}{17pc}
\includegraphics[width=17pc]{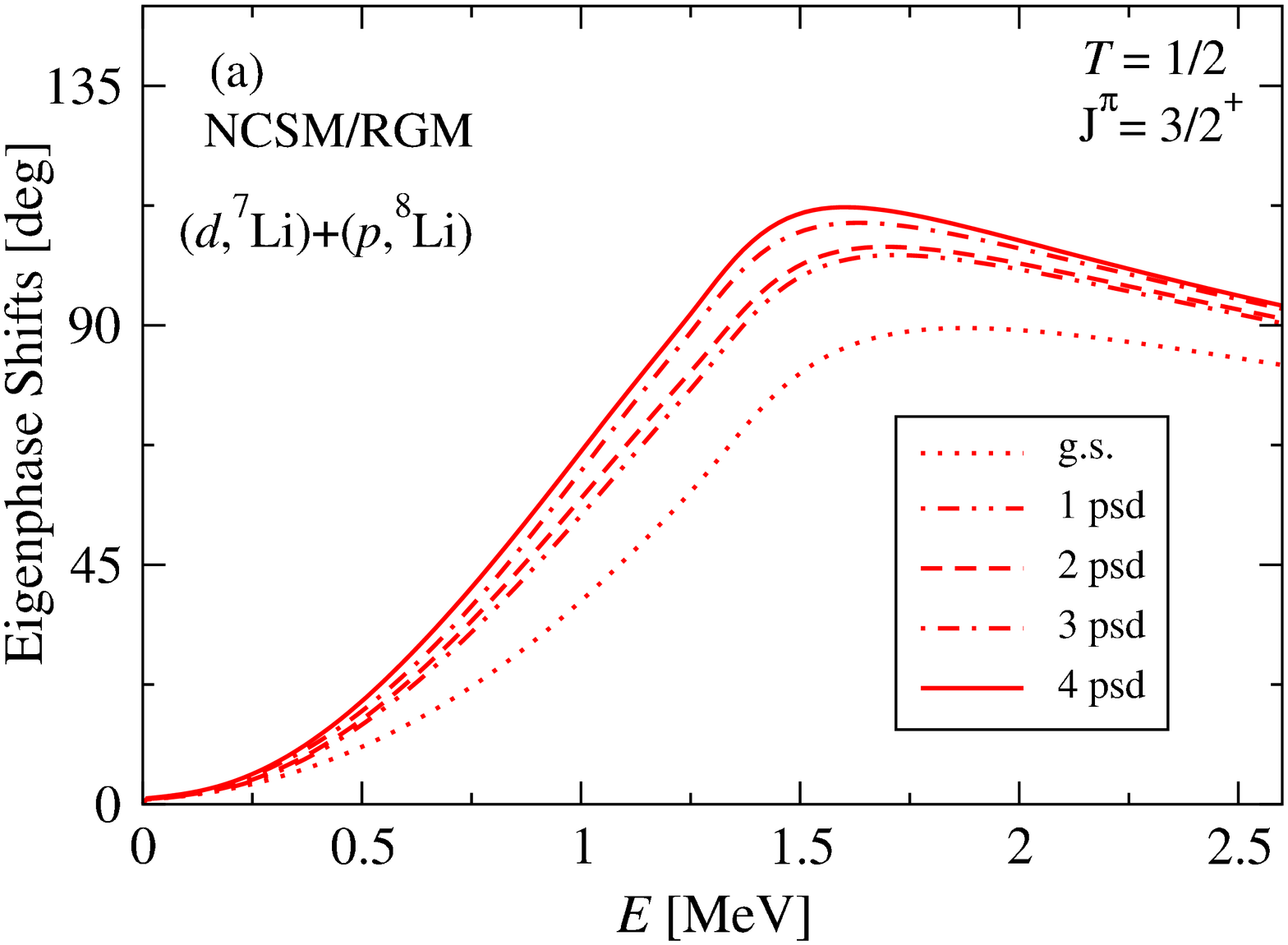}
\end{minipage}\hspace{3pc}%
\begin{minipage}{17pc}
\includegraphics[width=17pc]{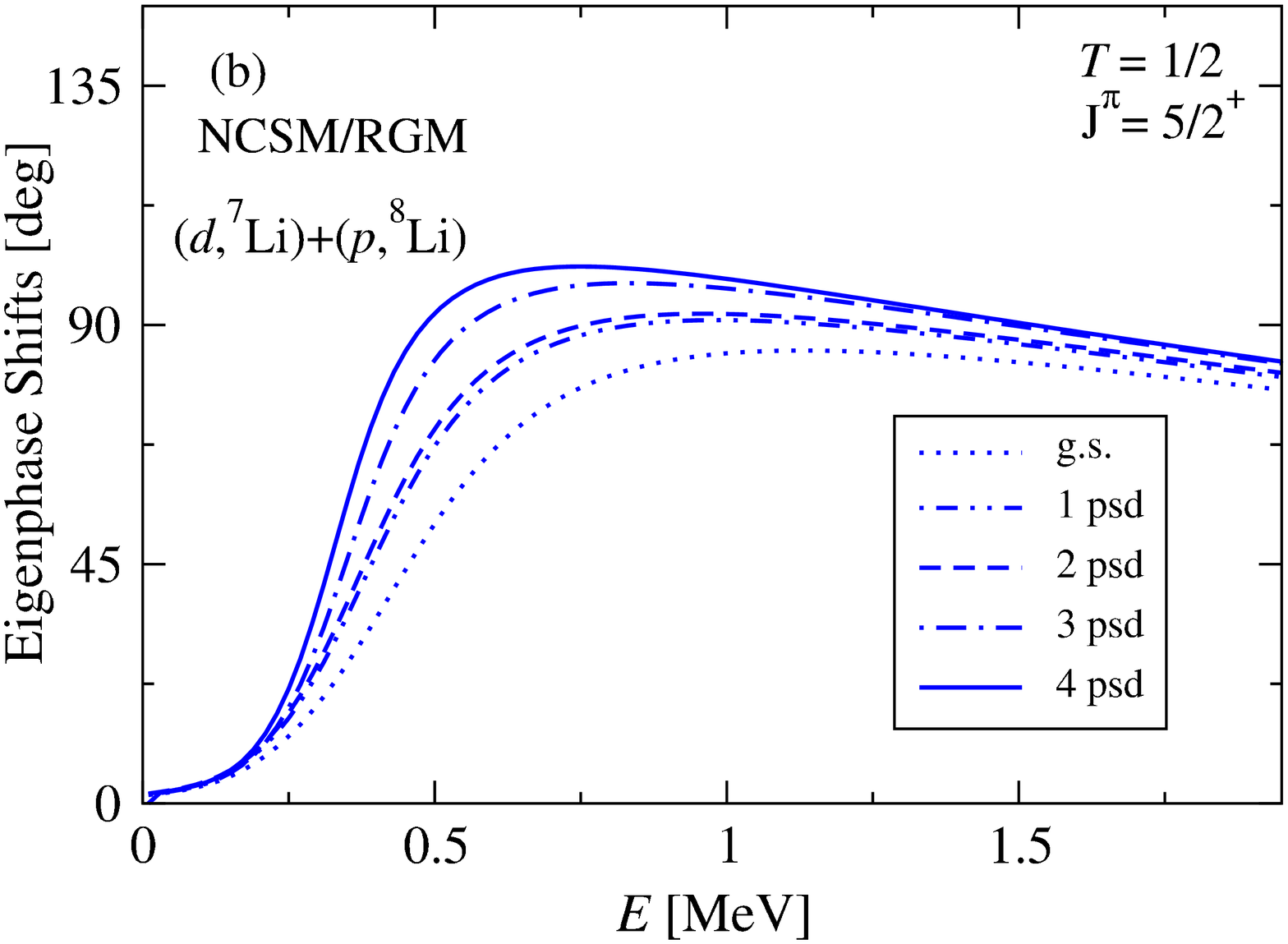}
\end{minipage} 
\caption{\label{eigen_plus_pseudo}Trend of convergence with respect to the deuteron pseudostates of Table~\ref{tab:d_pseudo}, for the calculated (a) $J^\pi T = \frac{3}{2}^+ \frac{1}{2}$ and (b)  $J^\pi T = \frac{5}{2}^+ \frac{1}{2}$ eigenphase shifts within the coupled ($d,^7$Li)+($p,^8$Li) NCSM/RGM basis, as a function of the relative kinetic energy in the c.m.\ frame and with respect to the $p$+$^8$Li threshold. The SRG-N$^3$LO NN potential with $\Lambda$=2.02 fm$^{-1}$, and the HO frequency of $\hbar\Omega$=20 MeV were used.}
\end{figure}

The positive-parity eigenphase shifts show a resonant shape just above  the $p$+$^8$Li threshold. In particular the dominant eigenphase shift carries the $\left(\frac{5}{2}^+ \frac{1}{2}\right)$ quantum numbers. The deformation and virtual breakup of the deuteron can be important even at low energies, therefore they may influence these resonances located just above the reaction threshold. In this respect, we analyse the influence of the deuteron pseudostates in Fig.~\ref{eigen_plus_pseudo}, where the dependence of the $J^\pi T = \frac{3}{2}^+ \frac{1}{2}$ and  $\frac{5}{2}^+ \frac{1}{2}$ eigenphase shifts on the number of deuteron states in the $^{3}S_1$-$^{3}D_1$ channel  is shown. Both resonances  are significantly enhanced by the inclusion of the deuteron pseudostates. We cannot claim that the solid line curves in   Fig.~\ref{eigen_plus_pseudo}, corresponding to 4 pseudostates included in the model space, are fully converged. However, considering the $\frac{3}{2}^+$ eigenphase shift in Fig.~\ref{eigen_plus_pseudo}(a), the relative difference between the resonance positions computed with 3 and 4 pseudostates is within 5\%, whereas the position of the resonance in the calculation without the deuteron continuum is shifted by 50\% with respect to our best attempt. Therefore, we are not expecting  significant changes of the shape of the eigenphase shifts, by adding more pseudostates in the calculation.

\begin{figure}[h]
\begin{minipage}{17pc}
\includegraphics[width=17pc]{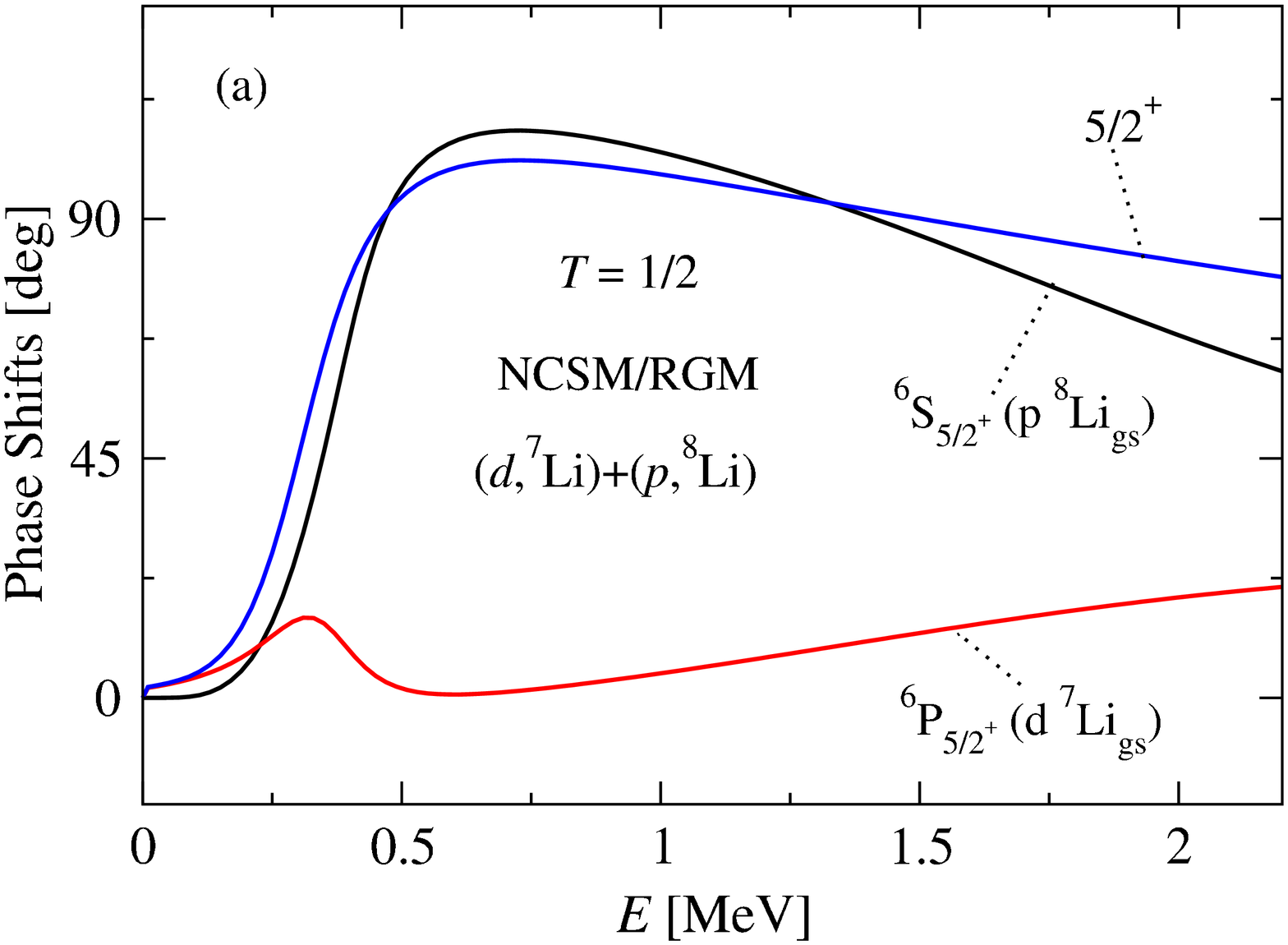}
\end{minipage}\hspace{3pc}%
\begin{minipage}{17pc}
\includegraphics[width=17pc]{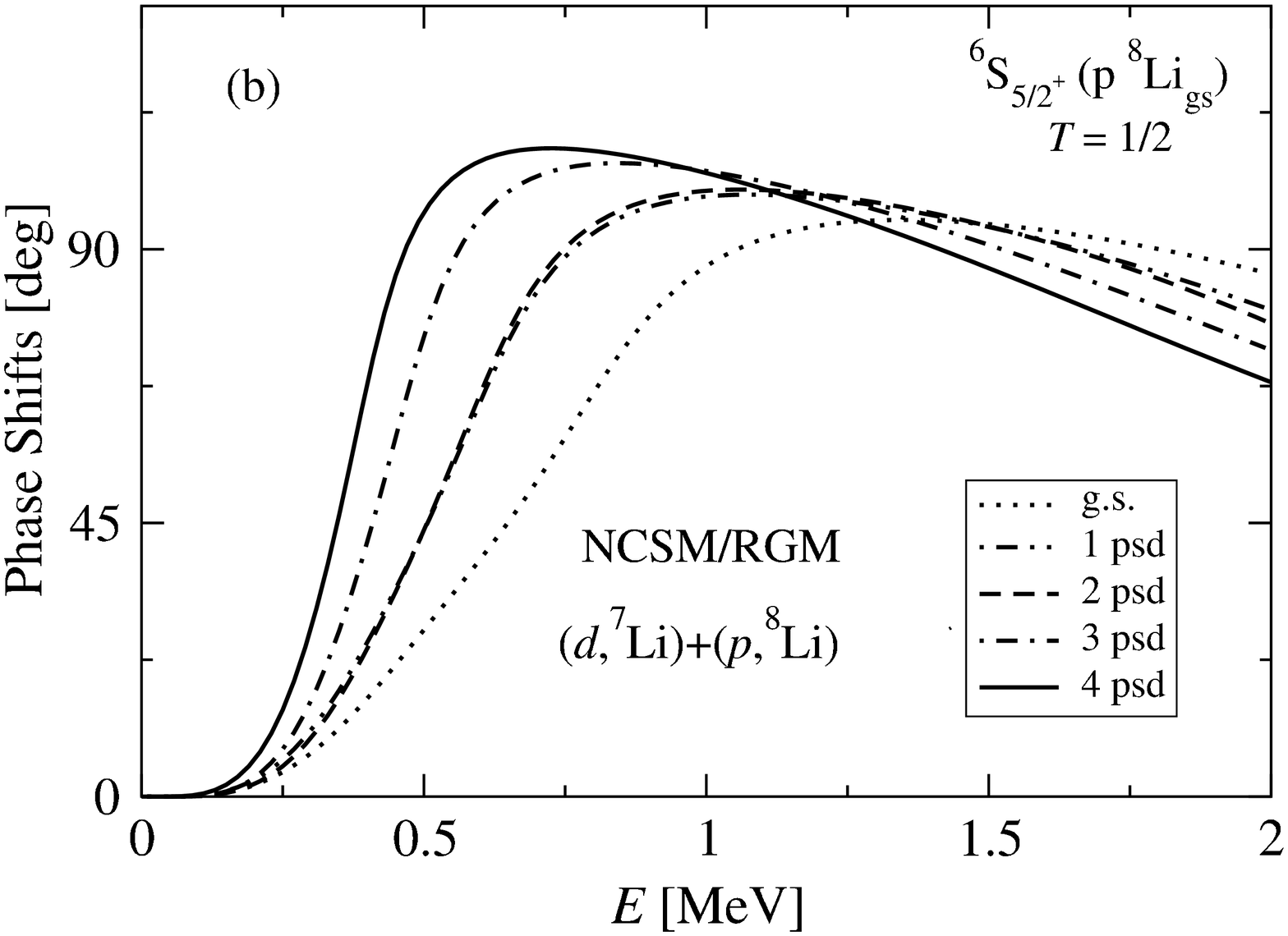}
\end{minipage} 
\caption{\label{phase_shift_5half} (a) Eigenphase shifts for $J^\pi T = \frac{5}{2}^+\, \frac12$  (solid blue line) compared to the $d$-$^7$Li (solid red line) and $p$-$^8$Li (solid black line) elastic phase shifts contributing to the same $J^\pi T$ scattering amplitude through a $P$- and $S$-wave respectively. (b) Trend of convergence with respect to the deuteron pseudostates of Table~\ref{tab:d_pseudo}, for  the $p$-$^8$Li $^6S_{5/2^+}$ elastic  phase shift.
All results 
were obtained  within the coupled ($d,^7$Li)+($p,^8$Li) NCSM/RGM basis, and 
are plotted as a function of the relative kinetic energy in the c.m.\ frame, with respect to the $p$+$^8$Li threshold.}
\end{figure}

Figure~\ref{phase_shift_5half}(a) further compares the two main phase shifts, labeled with the notation $^{2s}\ell_{J^{\pi}}$,  contributing to the $J^\pi T = \frac{5}{2}^+ \frac{1}{2}$ resonant state.  In the $d$-$^7$Li mass partition (solid red line), the $^7$Li g.s. is coupled to the deuteron in the relative $P$-wave motion, while the g.s. of $^8$Li has an $S$-wave coupling with the proton in the $p$-$^8$Li channel (black solid line), giving rise to a phase shift with a clear resonant behavior. The dependence of the $p$-$^8$Li $^6S_{5/2^+}$ channel on the number of deuteron pseudostates is shown in Fig.~\ref{phase_shift_5half}(b). Again, we see that the inclusion of the pseudostates in the model space  changes drastically the position of the resonance,  which is shifted by $\sim$600 KeV  towards the $p$+$^8$Li threshold  when 4 pseudostates are taken into account on top of the deuteron ground state.

%
 The non-resonant behavior of the channel states in the $d$-$^7$Li mass partition is a general feature of our calculations when both binary-cluster channels are coupled together. For instance, the $^6P_{5/2^+}$ channel, which is  a strong resonance in the uncoupled calculation with only $d$-$^7$Li mass partition, appears to be quenched in Fig.~\ref{phase_shift_5half}(a) (see Ref.~\cite{PhysRevC.93.054606} for a thorough discussion). 
 
\begin{figure}[h]
\begin{minipage}{17pc}
\includegraphics[width=17pc]{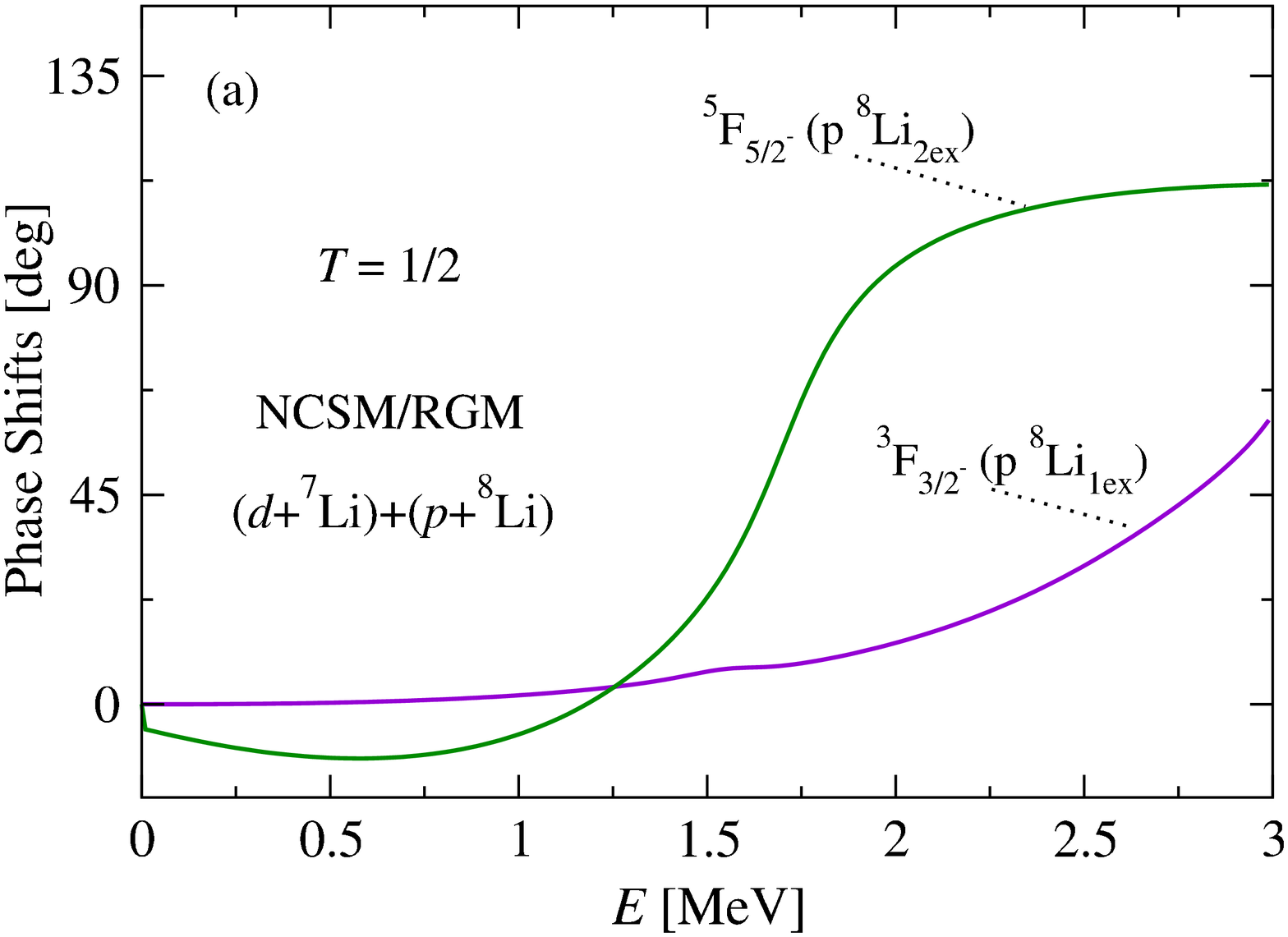}
\end{minipage}\hspace{3pc}%
\begin{minipage}{17pc}
\includegraphics[width=17pc]{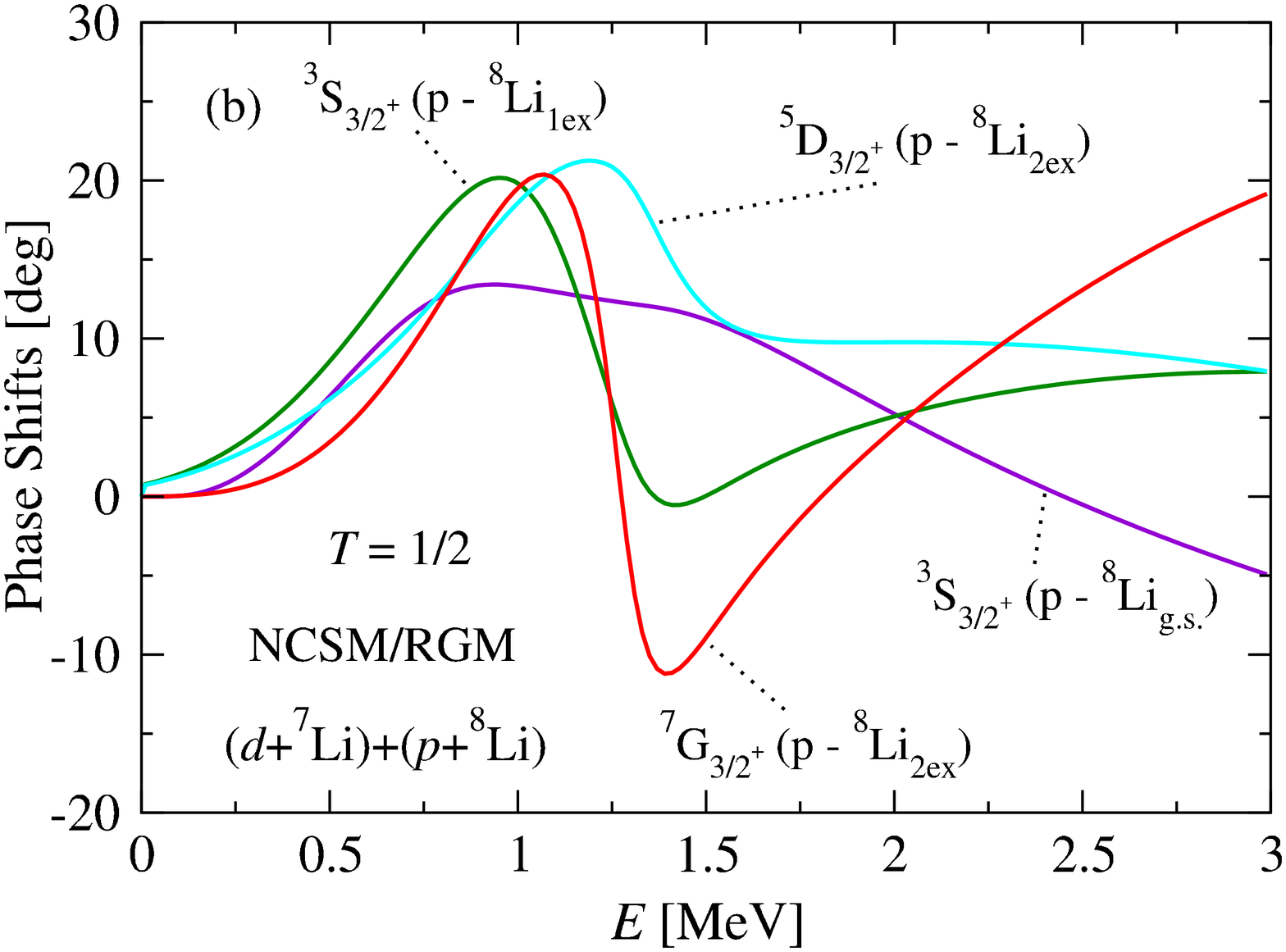}
\end{minipage} 
\caption{\label{phase_minus_plus}Calculated (a) negative and (b) positive-parity  phase shifts contributing to the partial waves shown in Figs.~\ref{eigen_minus_plus}, with the corresponding description of the binary-channel states in terms of the cluster state, total spin $s$   and relative-motion angular momentum $\ell$ of the binary-channel, according to the notation $^{2s}\ell_{J^{\pi}}$ (see Eq.~(\ref{eq:trial})).  The SRG-N$^3$LO NN potential with $\Lambda$=2.02 fm$^{-1}$, and the HO frequency of $\hbar\Omega$=20 MeV were used.}
\end{figure}

 In Figs.~\ref{phase_minus_plus}(a) and (b) we show the negative- and positive-parity phase shifts corresponding to the channel states  that contribute to the $J^\pi T$ states of Figs.~\ref{eigen_minus_plus}(a) and (b), respectively. They all belong to the $p$-$^8$Li mass partition and none of them, except for $^{5}F_{5/2^{-}}$, shows a clear resonant behavior.
%
We note that in the negative-parity partial waves, the dominant channels contain an excited state of $^8$Li; also the $J^\pi T=\frac{3}{2}^+\,\frac{1}{2}$ eigenphase shift contains contributions from the $^8$Li excited states, and consequently it is shifted at higher energy compared to the $J^\pi T=\frac{5}{2}^+\,\frac{1}{2}$ eigenphase shift. From this survey on the phase shifts, we conclude that the partial wave $\frac{5}{2}^+\,\frac{1}{2}$, which corresponds the first resonant state above the $p+^8$Li threshold, is the only one showing a significant contribute of the $d-^7$Li channels, which are otherwise suppressed in the calculations coupling both the mass partitions.

\subsection{\label{sec:cross_sec}${}^{7}$Li($d$,$p$)${}^{8}$Li integrated cross section}

The shape of the excitation functions of the ${}^{7}$Li($d$,$p$)${}^{8}$Li transfer reaction above the $d$+$^7$Li and $p$+$^8$Li thresholds, at about 16 MeV with respect to the ${}^{9}$Be g.s., reflects the pattern found in the (eigen)phase shifts of Figs.~\ref{eigen_minus_plus},~\ref{phase_shift_5half} and ~\ref{phase_minus_plus}.

\begin{figure}[h]
\begin{minipage}{17pc}
\includegraphics[width=17pc]{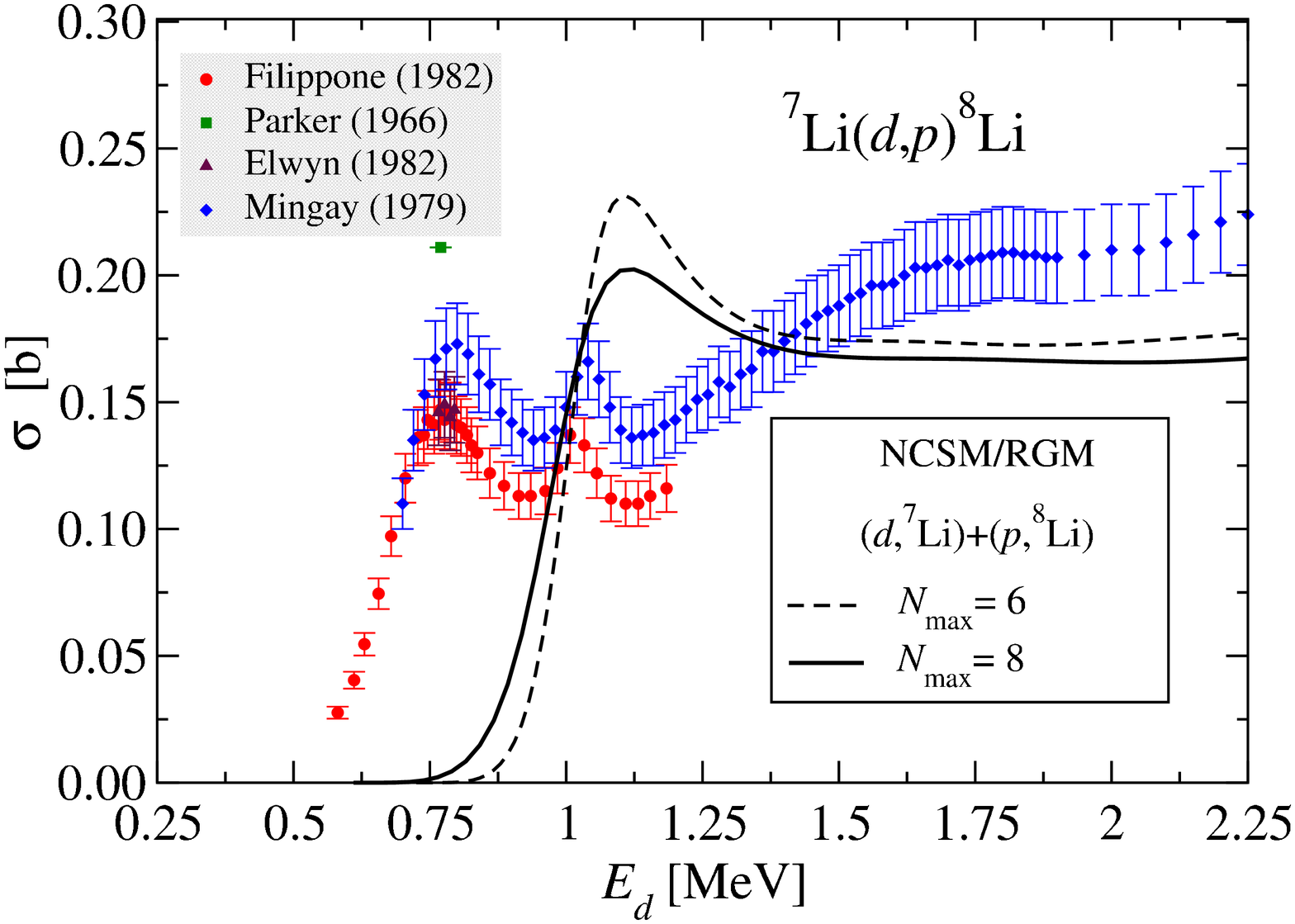}
\caption{\label{sigma_tot} $^7$Li($d$,$p$)$^8$Li integrated cross section for deuteron laboratory energies up to 2.25 MeV computed within the NCSM/RGM approach at $N_{\rm max}= 6$ (thin-dashed line) and 8 (solid line) compared to the experimental data from Refs.~\cite{Parker1966,Mingay1979, Elwyn1982,Filippone1982} (symbols).}
\end{minipage}\hspace{3pc}%
\begin{minipage}{17pc}
\includegraphics[width=17pc]{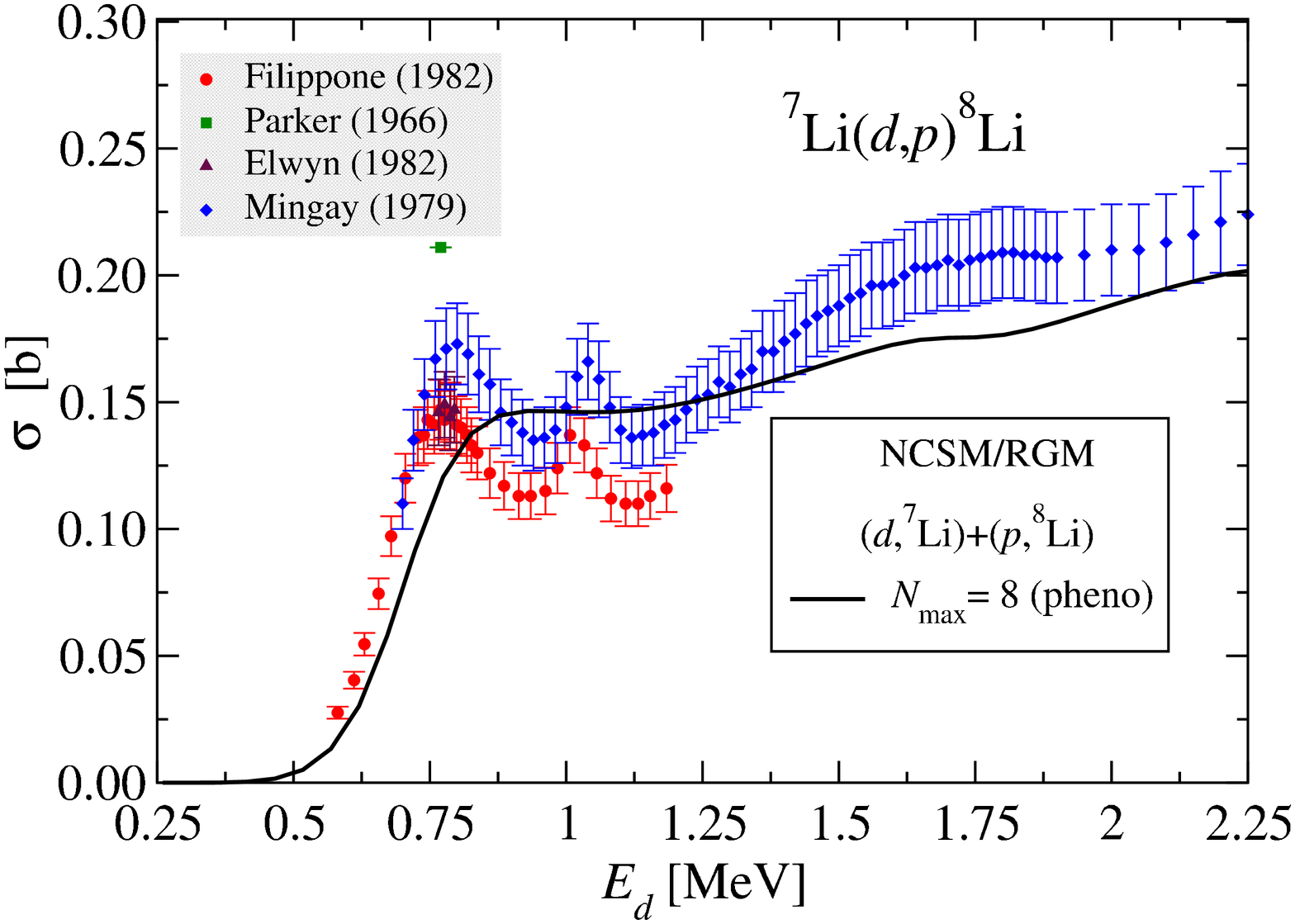}
\caption{\label{sigma_tot_pheno} Same as Fig.~\ref{sigma_tot}, but for the integrated cross section computed in the NCSM/RGM phenomenology approach, in which the experimental energy difference between $d+^7$Li and $p+^8$Li thresholds is taken as input in the calculation (see text for details).}
\end{minipage} 
\end{figure}

In Fig.~\ref{sigma_tot}, we compare the calculated ${}^{7}$Li($d$,$p$)${}^{8}$Li integrated cross section to the experimental data of Refs.~\cite{Parker1966,Mingay1979, Elwyn1982,Filippone1982} for deuteron energies in the laboratory frame up to about 2.3 MeV, that is the energy range below the breakup threshold of the deuteron. The experimental data displayed in Fig.~\ref{sigma_tot} are concentrated around the deuteron kinetic energy of 0.78 MeV, where the cross section exhibits a resonance of width $\Gamma\approx$ 0.2 MeV. The position of this peak,  with recommended value of 0.147$\pm$0.011 b, is measured routinely to calibrate the $^7$Be  targets used in experimental studies of the $^7$Be($p$,$\gamma$)$^8$B radiative
capture~\cite{Adelberger1998}.


The present computational limit with respect to the size of the HO basis, is given by $N_{\rm max}= 8$, corresponding to the solid line in Fig.~\ref{sigma_tot}. We observe that the first resonant peak
of the cross section approaches the experimental E $\sim$ 0.78 MeV resonance, when we increase the size of the HO basis from $N_{\rm max} = 6$ to 8.  Nevertheless, the position of this first peak is overestimated by about 0.33 MeV, which is likely due to the fact that our wave function is not yet converged. The fact that we overestimate the position of the first peak by about 0.33 MeV is related to the underestimation   of the Q-value of the reaction, as it is computed from the values of binding energies in Table~\ref{tab:tableNCSMstate}: The experimental Q-value is -0.192 MeV, whereas the energies of the g.s. in our calculation give
a Q-value of -0.556 and -0.465 MeV for $N_{\rm max}$= 6 and 8, respectively. The wrong threshold affects the NCSM/RGM calculation and its impact on the computed cross section can be seen in Fig.~\ref{sigma_tot_pheno}, where the NCSM energies for the $d$, $^7$Li and $^8$Li clusters are correct in order to reproduced the $d+^7$Li and $p+^8$Li thresholds with a desired level of accuracy. The values of the clusters energies used to obtain the curve in Fig.~\ref{sigma_tot_pheno} are shown in Table~\ref{tab:tableNCSMstate}. This way to proceed, denoted as \lq NCSM/RGM phenomenology\rq, brings the calculated total cross section in fairly good agreement with the measured one in Fig.~\ref{sigma_tot_pheno}, with the position of the first peak slightly overestimated and the trend of the cross section qualitatively reproduced, except the second peak at about 1 MeV above the $d+^7$Li threshold. The lack of this peak in the calculated cross section could be due to the missing $^8$Be($\alpha$-$\alpha$)-$n$ mass partition in the model space.

An interesting issue regarding the resonant peak at $\sim$ 0.78 MeV concerns the determination of its spin and parity. With the reasonable assumption that the first peak in our integrated cross section corresponds to the first experimental resonance, we can contribute to decide between conflicting spin-parity assignments derived from different phenomenological R-matrix analyses~\cite{Decharge1972,Friedland1971}. As illustrated in the eigenphase shifts of Fig.~\ref{eigen_minus_plus} and phase shifts of Fig.~\ref{phase_shift_5half}, our calculation supports a $\frac{5}{2}^+$ spin-parity assignment,  as suggested by the analysis in Ref.~\cite{Friedland1971}.  In general, by studying the contribution of the different partial waves to the total cross section (see Fig.~9 of Ref.~\cite{PhysRevC.93.054606}), we saw that the positive-parity phase shifts reach the maximum of their gradient at lower energies than the negative-parity ones.



\section{Conclusion}
We revisited the application of the NCSM/RGM approach to  the ${}^{7}$Li($d$,$p$)${}^{8}$Li transfer reaction~\cite{PhysRevC.93.054606}. This study is the first application of an \textit{ab initio}  method to the deuteron-induced transfer reaction with p-shell ($A>4$) targets. 

We found that the interplay between deuteron-$^7$Li and proton-$^8$Li channels explains some features of the $^9$Be spectrum in the energy region where the two thresholds are open. In this work, we strengthen the conclusion that the first resonant peak in the integrated cross section of the reaction, detected at deuteron energy of 0.78 MeV, contains a significant contribution from both the mass partition channels. The other important resonances are instead dominated by the proton-$^8$Li channels, with a corresponding quenching of the deuteron-$^7$Li ones. 

The study of the evolution of the phase and eigenphase shifts curves with respect to the number of the deuteron pseudostates, has shown the crucial effect of the continuum given by the deuteron polarization and its virtual breakup. While the trend of convergence in terms of the pseudostates suggests that the inclusion of more pseudostates could still affect quantitatively the position of the resonances, the main conclusions of our study are confirmed: in particular, the discussion on the experimental spin-parity assignments of the 0.78 MeV resonance, which is used
as a calibration for the target thickness in the proton-capture experiments on $^7$Be. We found that our calculations support a spin-parity assignment of $J^\pi=\frac{5}{2}^{+}$ for this resonance, suggesting a reaction mechanism dominated by the coupling of the $P$-wave $d-^7$Li incoming channel to the $S$-wave in the $p-^8$Li exit channel.

\vspace{2cm}

\bibliography{proceeding_TNPI2016_3}

\end{document}